\documentstyle[11pt,aaspp4]{article}
\begin{document}
\newcommand{\Lya}{Lyman~$\alpha$}
\newcommand{\Lyb}{Lyman~$\beta$}
\newcommand{\za}{$z_{\rm abs}$}
\newcommand{\ze}{$z_{\rm em}$}
\newcommand{\cmtwo}{cm$^{-2}$}
\newcommand{\nhi}{$N$(H$^0$)}
\newcommand{\nzn}{$N$(Zn$^+$)}
\newcommand{\ncr}{$N$(Cr$^+$)}
\newcommand{\degpoint}{\mbox{$^\circ\mskip-7.0mu.\,$}}
\newcommand{\halpha}{\mbox{H$\alpha$}}
\newcommand{\hbeta}{\mbox{H$\beta$}}
\newcommand{\hgamma}{\mbox{H$\gamma$}}
\newcommand{\kms}{\,km~s$^{-1}$}      
\newcommand{\minpoint}{\mbox{$'\mskip-4.7mu.\mskip0.8mu$}}
\newcommand{\mv}{\mbox{$m_{_V}$}}
\newcommand{\Mv}{\mbox{$M_{_V}$}}
\newcommand{\peryr}{\mbox{$\>\rm yr^{-1}$}}
\newcommand{\secpoint}{\mbox{$''\mskip-7.6mu.\,$}}
\newcommand{\sqdeg}{\mbox{${\rm deg}^2$}}
\newcommand{\squig}{\sim\!\!}
\newcommand{\subsun}{\mbox{$_{\twelvesy\odot}$}}
\newcommand{\et}{{\it et al.}~}
\def\h50{\, h_{50}^{-1}}
\def\hbl{km$^{-1}$~Mpc$^{-1}$}
\def\ltsima{$\; \buildrel < \over \sim \;$}
\def\simlt{\lower.5ex\hbox{\ltsima}}
\def\gtsima{$\; \buildrel > \over \sim \;$}
\def\simgt{\lower.5ex\hbox{\gtsima}}
\def\arcs{$''~$}
\def\arcm{$'~$}
\title{{\it HUBBLE SPACE TELESCOPE} IMAGING OF STAR-FORMING GALAXIES AT 
REDSHIFTS $Z > 3$\altaffilmark{1}}
\author{\sc Mauro Giavalisco\altaffilmark{2}}
\affil{Observatories of the Carnegie Institution of Washington, 813 Santa
Barbara Street, Pasadena, CA 91101}
\affil{e-mail: mauro@ociw.edu}
\author{\sc Charles C. Steidel\altaffilmark{3,4}}
\affil{Palomar Observatory, California Institute of Technology, Mail
Stop 105-24, Pasadena, CA 91125}
\affil{e-mail: ccs@astro.caltech.edu}
\author{\sc F. Duccio Macchetto\altaffilmark{5}}
\affil{Space Telescope Science Institute, 3700 San Martin Dr. Baltimore, MD 
21218}
\affil{e-mail: macchetto@stsci.edu}

\altaffiltext{1}{Based on obervations with the NASA/ESA {\it Hubble Space 
Telescope} obtained at the Space Telescope Science Institute which is operated 
by AURA under NASA contract NAS 5-26555.}
\altaffiltext{2}{Hubble Fellow.}
\altaffiltext{3}{Alfred P. Sloan Foundation Fellow.}
\altaffiltext{4}{NSF Young Investigator.}
\altaffiltext{5}{Affiliated with the Space Science Department, ESA.}

\begin{abstract}
We present {\it Hubble Space Telescope} images of star-forming galaxies 
at redshifts $z>3$. These galaxies have been selected using ground--based 
images and color criteria 
sensitive to the presence of a Lyman discontinuity 
in the otherwise flat (in $f_{\nu}$ units) UV spectral energy distribution of 
unreddened star formation. The spectroscopic confirmation of these $z > 3$ 
galaxies is reported in a companion paper (Steidel et al. 1996). The {\it HST}
images, which probe the rest-frame UV between 1400 and 1900 \AA, show that the
morphologies of the $z>3$ galaxies are generally compact and exhibit a relatively high
degree of spherical symmetry, although we find a few cases of more diffuse light
profiles and several cases where the objects are comprised of multiple compact   
structures. Overall, the dispersion of morphological properties 
is relatively narrow, in contrast to the variety found in star-forming 
galaxies at intermediate redshifts ($z\sim 1$). The galaxies with compact
morphology are typically characterized by a small but resolved ``core'', 
approximately $\simlt 0.7$ arcsec in radius, or about $5\h50$ ($8.5\h50$)
kpc with $q_0=0.5$ (0.05), and half-light radii of 0.2--0.3 arcsec, or 
$1.4$--$2.1\h50$ ($2.4$--$3.6\h50$) kpc. These sizes and scale lengths
are similar to those of present-day bulges or intermediate-luminosity spheroids. 
The ``cores''  are often surrounded by lower surface-brightness  nebulosities,
generally asymmetrically distributed. The minority of more diffuse galaxies do
not possess this core, and an exponential function provides a very good fit to
their light profiles. In contrast to highly elongated or irregular structures, 
such as ``chain galaxies'', that are found at $z \sim 1$, the $z>3$ galaxies are 
characterized by a relatively high degree of spherical symmetry. The morphological 
properties, space density, star-formation rates, masses, and early epoch of 
the star-formation phase all support the hypothesis that we 
have identified the progenitors of present-day luminous galaxies at the epoch
when they were forming the stars of their spheroidal components. 
\end{abstract}

\section{INTRODUCTION}

The physics of galaxy formation remains largely unconstrained from an empirical
perspective. In the last few years, several deep redshift surveys (Lilly et
al. 1995a and 1995b; Steidel et al. 1995; Glazebrook et al. 1995a; Cowie et
al. 1994; Cowie, Hu \& Songaila 1995a) and deep post-refurbishment {\it HST} 
imaging (Driver et al. 1995a and 1995b; Glazebrook et al. 1995b; Schade et
al. 1995; Cowie Hu \& Songaila 1995b) have extensively probed the 
evolutionary state of galaxies at intermediate redshifts ($z<1.0$),
corresponding to $<55$ ($<60$)\% of the life of the Universe ($H_0=50$
\hbl\ and $q_0=0.5$ ($0.05$). This work seems to show that the evolution 
of the luminosity function has followed rather diverse tracks for galaxies 
of different luminosity and morphological type. A common conclusion is that the 
population of luminous galaxies, i.e., the systems currently identified as
relatively massive ellipticals and spirals, are characterized by at most a
modest amount of evolution in luminosity and/or number density since $z \sim 1$. 
The available data therefore 
suggests that the epoch of formation of the most massive and oldest
systems predates that probed by the current surveys. 
While there has evidently been substantial evolution of later type
systems in number and/or luminosity in the relatively recent past, the population
of galaxies possessing a substantial spheroidal component (i.e., ellipticals
and early--type spirals) has been remarkably quiescent over the redshift 
range probed by the redshift surveys, suggesting that the important epoch for 
their formation lies far beyond $z \sim 1$. The fact that spheroidal systems, 
which contain approximately half of the present-day stars (Schechter \&
Dressler 1987), had assembled rather early in the course of the evolution 
has long been the rationale for searches for ``primeval'' galaxies. 
Establishing the epoch and the formation mechanism of
these systems will place significant constraints on theories of
galaxy and structure formation.  

In a companion paper (Steidel et al. 1996, S\&96 hereafter) 
we report the discovery of a 
substantial population of star-forming, but otherwise normal (i.e. {\it non--
AGN}), galaxies at redshifts $z>3$. These galaxies were found using color 
criteria sensitive to the presence of a Lyman discontinuity (due to a combination
of a galaxy's opacity to its own UV continuum radiation, the intrinsic energy
distribution of hot stars, and the opacity of intervening gas at high redshift)
in the otherwise flat (in $f_{\nu}$ units) and featureless 
spectral energy distribution (SED) of unreddened star formation (see Steidel,
Pettini, \& Hamilton 1995). 

The color criteria are very efficient in selecting galaxies at high redshifts,
and although we have been able only recently to confirm them spectroscopically,
over the last few years we have been collecting a fairly large sample of $z>3$
galaxy candidates (Steidel \& Hamilton 1992, 1993; Steidel et al. 1995; 
Giavalisco et al. 1996, in preparation) and we have been investigating their 
morphological properties with the {\it Hubble Space Telescope} ({\it HST}) 
(see, e.g. Giavalisco et al. 1995). At the time of this writing, only 23 
galaxies out of about 100 have securely measured redshifts. However, as we 
have detailed in S\&96, we expect that a very high fraction, probably 
$\simgt 90$\%, of the candidates not yet spectroscopically confirmed are 
also in our targeted redshift range of $3.0 \le z \le 3.5$. We have already 
presented a number of arguments linking the population we have identified 
with the luminous galaxies of the present epoch (S\&96). In this paper,
we present deep {\it HST} images of 19 of the ``Lyman break'' galaxies, 6 of 
which have secure redshift measurements in the range $2.8 \simlt z \simlt 3.4$.  
The images probe the rest-frame UV spectrum in the range $1400$--$1900$ \AA\ 
and have resolution high enough that we can attempt a quantitative discussion
of their morphological properties. Thus, for the first time, we can
characterize in a statistically significant way the evolutionary status of
galaxies at a time when the universe was $\simlt 20$\% of its current age. 

\section{THE DATA}

The data set consists of images obtained with {\it HST} and WFPC2 of 19 galaxies
from 3 fields, which we designate by their approximate coordinates, 0000-263, 
0347-383, and 2217-003, respectively, for a total of 7 different {\it HST} 
pointings, with 4 independent pointings in the 22 hour field. The two fields 
0000-263 and 0347-383, the former imaged through the F606W and F702W passbands,
the latter through the F702W only, were observed by us to follow-up the 
ground-based detections of the $z>3$ galaxies and galaxy candidates presented 
by Steidel \& Hamilton (1992), Giavalisco, Steidel \& Szalay (1994), and
Steidel, Pettini \& Hamilton (1995). For the remaining 4 fields, which have
all been observed through the F814W passband, we have followed the inverse 
strategy of performing our ground--based photometric searches for $z>3$
galaxies in existing deep {\it HST} fields to improve the overall observing 
efficiency. The case of the 2217-003 field is particularly fortuitous, as 4 
different {\it HST}
pointings are available for it, all of which could be contained within a single 
pointing of the COSMIC camera at the prime focus of the Palomar 5m telescope 
during the primary ground-based search. In this field, called SSA22 in their 
nomenclature, Cowie et al. (1995b) obtained one {\it HST} pointing to
follow-up their $K$-band selected galaxy survey, while Lilly et al. (1995,
{\it HST} Cycle-5 observations, private communication) obtained 3 more
pointings to follow-up their Canada-France Redshift Survey. We have identified
2 $z>3$ candidates, one spectroscopically confirmed, in the SSA22 field,
and 3 in the Lilly \et fields, a total of 4 of which are presented in this 
work (one of the Lilly \et fields was not available at the time of this writing). 
The details of the {\it HST} exposures used in the present work are summarized in
Table 1. 

In the 0000-263 field, because the spectral energy distribution (SED) of the 
$z>3$ galaxies is essentially flat (in 
$f_{\nu}$ units) and the images in the 2 passbands have comparable S/N,
we have registered and added them together to obtain a deeper
exposure, with appropriate adjustments to the photometric zero point. 
Although the equivalent band width of this combined image is 
$\approx 2060$ \AA, the rest-frame portion of the spectrum which it
probes is $485$ \AA\ wide, only $\sim 40$\% wider than the B1 passband
adopted for the UIT camera in the Astro-1 and Astro-2 missions (Stecher et 
al. 1992). With the exception 
of galaxy 0000-263-C10, observed only in the F606W frame, all of the 
galaxies in the 0000-263 field are included in the combined frame used for the
morphological analysis.

Table 2 presents the sample of galaxies observed (for convenience we used 
the same nomenclature as in S\&96), detailing redshift (when available), 
passband, aperture and isophotal magnitudes in the $AB$ system, star-formation
rates (when redshift is available) computed from the isophotal magnitudes 
using the ``continuous star formation models'' by Leitherer, Robert, \&
Heckman (1995) with a Salpeter initial mass function and upper mass cut-off of
$80$ M$_\odot$ (see also discussion in S\&96), and morphological parameters (see
discussion below), including measures of the axial ratios in the last column. 
As it can be noted, in a number of cases the magnitudes in the $r=0.7$ arcsec 
aperture are actually slightly brighter than the isophotal magnitudes, in all 
cases by an amount smaller than $0.1$ mag. This is due to the errors
associated with the magnitude measurements at these flux levels, and to the 
compactness of the $z>3$ galaxies, as we discuss in the next section.

\section {THE MORPHOLOGY OF GALAXIES AT $Z>3$}

Figure 1 shows a mosaic of the galaxies. Figure 2 presents a mosaic of their
radial profiles. From a visual inspection of the two figures it is clear that 
the majority of the $z>3$ galaxies have compact morphologies, typically 
characterized by a bright ``core'',  often surrounded by more diffuse 
nebulosities with significantly lower surface brightness. The core is
typically $\simlt 0.7$ arcsec in radius, with a characteristic central surface
brightness $SB_c\sim 23$ mag arcsec$^{-2}$. The nebulosities clearly deviate
from the core profile. 
However, there are 4 cases of more diffuse galaxies in the sample,
C13, C27, and C28 in the 0000$-$263 field (we note that C28 is a candidate
damped Lyman $\alpha$ absorber in the spectrum of Q0000$-$263--see Steidel \&
Hamilton 1992), and C24 in the SSA22-L3 field. The diffuse galaxies have
larger sizes than the compact galaxies, do not possess an obvious ``core'' and
have lower central surface brightness. There are also 2 cases of
``double-core'' galaxies, SSA22-CW-C12 and 0000-263-C12 (no radial profile has
been plotted for this last galaxy, given the very close proximity of the two 
``cores'' and the difficulty in defining a central light concentration). 
Finally, the galaxy 0000-263-C14, appears to be an interacting system. 

We have quantitatively examined the morphology of the galaxies in two ways. To 
analyze their radial light profile we fit elliptical isophotes to 
the {\it HST} images. In principle, one should fit the center, ellipticity 
and position angle of each individual isophote. However, the presence of the 
substructures and the relatively small number of pixels that comprise the image 
make this method inaccurate. We adopted a procedure wherein we used the center
derived from the inner isophote (typically $r<0.3''$); keeping this center fixed, 
we selected whatever combination of ellipticities and position angle best fit the
radial light profiles. In practice, we fit each radial profile with
both an $r^{1/4}$ and an exponential law to determine which of the two best
describes each galaxy. 

These fits are intended to broadly classify the light profiles and are
not meant to imply that a particular galaxy rigourously follows a given model.
In addition, we caution that we are fitting functions designed to model the 
light distribution of present-day galaxies at $4500$ \AA\ rest-frame, and
characterized by a very modest amount of star formation, to galaxies at $z>3$ 
observed at $1600$ \AA\ rest-frame and with a star-formation rate one order of
magnitude (or more) higher. Clearly, we do not yet know whether the same 
physical interpretations relating morphology to the dynamical state of the
galaxies would hold. In Figure 2, we show the fitted $r^{1/4}$ (continuous) 
and exponential (dotted) laws, while in Table 2 we report which of the two 
functions provides a better fit, together with the scale length ($r_e$ for the 
former, $r_0$ for the latter). 

Because of the approximate nature of the analysis above, we have not
derived measures of the light concentration from the parameters of the fit,
as deviations from the adopted model and inaccurate modelling of the wings 
of the light distribution would have
resulted in large errors. Rather, we have measured the isophotal magnitude
(using FOCAS) and a set of concentric aperture magnitudes, centered at the
peak of the light distribution (the centering box had a size of 5 pixels), 
to produce a growth curve and derive the half-light radius. These magnitudes 
(including the aperture sizes), and the half-light radii (one relative to the 
''core'', defined to have $r<0.7$ arcsec, and one to the whole galaxy) are 
collected in Table 2. 

Only 6 galaxies out of 19 in the {\it HST} sample have rigorously measured 
redshifts (see S\&96 for the complete list of secure redshifts). 
However, the efficiency of the color selection is very high, with about 
$70$\% of those attempted spectroscopically being confirmed, and with the 
remaining $30$\% indeterminate but still consistent with the same range of 
redshifts (there are {\it no cases} of ``interlopers'' observed so far, i.e., 
galaxies with redshifts less than the targeted range). In the following 
discussion we have assumed that all 19 of the galaxies are indeed at high 
redshift. 

Finally, two galaxies (0000-263-C14 and SSA22-CW-C12) 
clearly show a structure in the form of double nuclei. In these cases 
it is not possible to compute a unique position for the centroid of 
the light distribution; therefore, we have carried out the above analysis 
for the two sub-components separately. In Table 2 these objects are labeled 
with an ``a'' and a ``b'' suffix following the name of the galaxy. In each 
case, the 2-D spectrograms showed that the two sub-components of the 
galaxies are both placed at the same redshift. 

\section {DISCUSSION AND CONCLUSIONS}

Ideally, one would like to understand the nature of the $z>3$ galaxies and
their place in a general scheme of galaxy evolution. Although this will 
certainly require additional work, several interesting results have already 
emerged from the analysis of the {\it HST} images:

\noindent{\bf 1)} Most of the $z>3$ galaxies are characterized by a compact 
morphology, generally having one central ``lump'' where the majority of the 
light is concentrated. The size of the central concentration is typically  
$\simlt 1.5$ arcsec in 
diameter, which at redshift $z=3.25$ (the mean value for our survey) 
corresponds to $10.5\h50$ ($18\h50$) kpc. In the following, we refer 
to this central lump as the ``core''. The core typically contains about 
$90$--$95$\% of the total luminosity of the galaxy, and has a half-light
radius in the range 0.2--0.3 arcsec, corresponding to $1.4$--$2.1\h50$ 
($2.4$--$3.6\h50$) kpc. Since at the observed rest-frame far--UV wavelengths 
the emission is directly proportional to the formation rate of massive stars, 
one can conclude that $\sim 90$--$95$\% of the stars that are being formed 
in these galaxies are concentrated in a region whose size is that of a 
present-day luminous galaxy. Moreover, the characteristic central 
concentration of the star formation has a scale size similar to a 
present--day spheroid. If the morphology of the massive stars is
a good tracer of the overall stellar distribution, then the light distribution
of the core is consistent with a dynamically relaxed structure. 

\noindent{\bf 2)} The core is very often surrounded by low surface
brightness nebulosities, generally distributed asymmetrically, and
which may extend for a few arcsecs (see, e.g. 0000-263-C09 and 0347-383-N05). 
These nebulosities are clearly observed in Figure 2 as deviations from the 
more regular profile of the core. We note that the presence of such halos is
consistent with the intense star-formation activity observed in the $z>3$
galaxies, even if the morphology of the underlying stellar distribution is 
relatively regular. For instance, the presence of extended gaseous 
components is expected from the intense supernovae rate that must apply 
in these systems (Ikeuchi \& Norman 1991). Also, halos with irregular 
morphology and lower $SFR$ are consistent with the dissipative collapse 
(Baron \& White 1987) of the cores. In such a scenario a 
core-halo segregation is actually expected due to the increased cloud 
collision rate and $SFR$ in the denser, more collapsed regions (Silk \& Norman 
1981). 

\noindent{\bf 3)} The central SB of the compact $z>3$ galaxies is 
consistently $\sim $23 mag arcsec$^{-2}$. At the observed rest-frame
wavelengths,  the surface brightness is proportional to the star-formation
efficiency. Given that these galaxies all have comparably small UV extinction, 
we can conclude that they have also comparable star-formation efficiency, possibly 
indicative of the interplay of similar physical processes in the core regions.

\noindent{\bf 4)} There are 4 cases of diffuse galaxies, namely 0000-263-C13, 
0000-263-C27, 0000-263-C28 and SSA22-L3-C24, whose light profiles are very 
well fitted by exponential laws. The central surface brightness of these 
galaxies is significantly lower than that of their more compact counterparts, 
showing an overall reduced star-formation efficiency. Interestingly, galaxy 
0000-263-C28 is a candidate to be one of the 2 known damped Lyman $\alpha$ systems 
in the spectrum of Q0000-263, either at $z=3.052$ or $z=3.390$. 

\noindent{\bf 5)} In 3 cases out of 19, SSA22-CW-C12, 0000-263-C12, and 
0000-263-C14, we observe galaxies with a multiple morphology, where two 
major light concentrations with similar apparent luminosity, two ``cores'' 
in our terminology, are separated by about 1 arcsec or less, corresponding to
$\simlt 7\h50$ ($12\h50$) kpc. The two individual sub-components are spatially 
resolved in the first case, where one is much more concentrated
than the other, barely resolved in the second case, and too small to conclude 
anything in the third case. In all cases, the galaxies clearly show extended 
diffuse nebulosity around or extending from them, which is suggestive of 
systems in interaction. This could be interpreted as evidence of hierarchical 
merging of sub-units into more massive systems, taking place with time scales 
about an order of magnitude shorter than the time span probed by the redshift
range $3.5 \ge z \ge 3.5$, 
or $t_{merg}\approx 5(8)\times$ $10^7$ yr. We note that in all cases the 
``merging'' units have smaller luminosity than the other systems, but
comparable central surface brightness. This is fully 
consistent with the interpretation of these galaxies as young spheroidal 
systems. Interestingly, both parents and daughters of this possible merging
scenario have comparable morphologies. 

\noindent{\bf 6)} In the sample available to us at the
present time, the geometry of the cores is characterized by a relatively 
high degree of spherical symmetry, and there are no cases of highly elongated 
structures, such as the ``chain galaxies'' presented by Cowie et al. (1995b). 
These chain galaxies are apparently formed by strings of star-forming regions 
of similar surface brightness. 
Given their knotty structure and the fact that they seem
to have comparable star-formation rates to those of the $z>3$ galaxies, 
if placed at $z>3$, we would expect that the ``morphological 
k-correction'' (i.e., the change in the apparent morphology of a galaxy due 
to the fact that the observed wavelength is shifted into the rest--frame far UV at high
redshifts) 
would make them appear even more 
elongated because of $(1+z)^{4}$ surface-brightness dimming of the more
diffuse regions that surround the knots. 
None among the 19 galaxies observed 
so far has, even approximately, such a morphology. We have quantified this by 
measuring the axial ratios of the isophotes, which provide an upper limit to the
axial ratios of the galaxy core regions. These values, reported in Table 2,
have a mean of $1.7$, and are larger than $\approx 1.5$--$2$ only for the markedly 
exponential galaxies (and are never larger than $\approx 3$). In comparison, 
Cowie et al. report unconvolved axial ratios as high as 9.5, with a mean value
of 4.7 for the chain galaxies. 
In general, the $z>3$ galaxies 
exhibit a relatively small dispersion in 
morphology, in contrast to the larger variety observed in the galaxies 
harboring most of the star formation at later epochs (cf. Cowie et al. 
1995a and 1995b; Driver et al. 1995a and 1995b; Glazebrook et al. 1995b)

In interpreting the 
far--UV morphologies of galaxies at substantial redshifts, one must bear 
in mind the strong surface brightness selection effects ($SB\propto (1+z)^{-4})$. 
On the other hand, the SED of unreddened star-forming galaxies is essentially 
flat from $\sim 1200$ \AA\ to $\sim 4000$ \AA\ rest-frame, and to even longer
wavelengths if 
the galaxy is young. At the wavelength of a typical WFPC2 filter (e.g. the
F702W with $\lambda_e\sim 7000$ \AA), this corresponds to observing galaxies 
in the redshift interval $0.75<z<4.4$ with minimal ``morphological 
k-corrections'' (see Giavalisco et al. 1996). This means that it is not unfair
to compare UV morphologies of the $z >3$ galaxies with those of the $z \sim 1$ 
systems. 

Although it is not yet clear which physical parameter(s) is responsible for
the variety of observed morphologies at intermediate redshift, the $z>3$ 
galaxies exhibit a high degree of morphological consistency, lending credence 
to the idea that we are observing a well defined ``population''. In this respect,
we note that the $z > 3$ galaxies were selected using ground--based images in 
which the degree of central concentration was completely unknown {\it a priori}
due to seeing effects (the ground based images have typical FWHM 
$\sim 0.9-1.3$ arcsec); thus we believe that the sample we have presented 
fairly represents the population of objects having the most substantial star 
formation rates at $3.0 \le z \le 3.5$. One could speculate that the 
morphological peculiarities observed at intermediate redshifts are caused 
in part by the triggering of substantial star formation episodes in 
galaxies with masses that are too small to remain intact in the face of their 
own star formation (e.g. Dekel \& Silk 1986). The $z >3$ objects, which have 
star formation rates as high as the starbursting galaxies at $z \simgt 1$ 
observed by Cowie et al. (1995a), clearly have enough binding energy to remain
relatively ``organized'' despite the very high supernova rate that would be 
predicted ($\sim 1$ yr$^{-1}$), given the number of short--lived O stars known
to be present.   

As shown in S\&96, the observed volume density of the $z>3$ 
galaxies is within a factor of a few
that of present--day galaxies with $L>L^*$, and their integrated
star-formation rate is at least 25\% of what is observed in the local universe. 
Rough dynamical estimates suggest that the galaxies are massive systems. In
view of the fact that galactic spheroids {\it must} have formed relatively
early to attain a state of quiescent evolution by $z \sim 1$, we find
the centrally concentrated star formation that characterizes the $z>3$
population, together with all the other established properties, compelling 
evidence that 
we are observing directly the ongoing formation of the spheroid components of
what would become the luminous galaxies of the present epoch. At $3\simlt
z\simlt 3.5$ we are probably seeing an epoch where the star formation was
concentrated primarily in the central regions of massive galaxies (the
``spheroid epoch''). This star formation appears to have ``migrated''
over time to more morphologically peculiar objects, and finally to
the present where the bulk of the star formation is distributed in spiral
disks. 

In any event, the $z>3$ galaxies we have identified are the sites of the
most active star formation at their epoch, and as a consequence they 
represent an important phase in the early history of galaxy formation. 
The properties of these objects must be reproduced by any theory attempting 
to explain the formation of normal galaxies. 

\acknowledgements

We would like to express our gratitude to Simon Lilly and the CFRS team for 
having allowed us to use their Cycle 5 imaging data in the SSA22 field prior 
to publication. It is a pleasure to thank Max Pettini for the several highly 
illuminating discussions. We also thank an anonymous referee for his/her very
useful comments on the manuscript. This research was supported by NASA grant
GO-05964.01-95A. MG acknowledges support from the Hubble Fellowship program 
through grant 
number HF-01071.01-94A, awarded by the Space Telescope Science Institute, 
which is operated by the Association of Universities for Research in 
Astronomy, Inc. under NASA contract NAS5-26555. CCS acknowledges support
from the Sloan Foundation and from the NSF through grant AST--9457446.

\newpage
\begin{deluxetable}{llccc}
\tablewidth{0pc}
\footnotesize
\tablecaption{The WFPC2 fields}
\tablehead{
\colhead{\#} & 
\colhead{Field} & 
\colhead{Passband} & 
\colhead{Exp. Time\tablenotemark{a}} & 
\colhead{SB limit\tablenotemark{b}} } 
\startdata
1  &  0000-263                   &     F606W     & 15600 & 29.31 \nl
2  &  0000-263                   &     F702W     & 27400 & 28.91 \nl
3  &  0000-263                   & F606W + F702W & 43000 & 29.73 \nl
4  &  0347-383                   &     F702W     & 18000 & 28.82 \nl
5  &  2217-003\tablenotemark{c}  &     F814W     &  6700 & 27.61 \nl
6  &  2217-003\tablenotemark{d}  &     F814W     & 28800 & 29.38 \nl
\enddata
\tablenotetext{a}{Total exposure time in seconds}
\tablenotetext{b}{$1$-$\sigma$ surface brightness fluctuations in AB mag
arcsec$^{-2}$ in the WF-3 chip}
\tablenotetext{c}{SSA22 pointing by Lilly et al. (1995), referred here as SSA22-L3}
\tablenotetext{d}{SSA22 pointing by Cowie et al. (1995), referred here as SSA22-CW}
\end{deluxetable}
\newpage
\begin{deluxetable}{lcccccccccccc}
\tablewidth{0pc}
\scriptsize
\tablecaption{The sample of $z>3$ galaxies}
\tablehead{
\colhead{\#} & 
\colhead{Name} & 
\colhead{$z$} &
\colhead{Filter} & 
\colhead{$m_{0.2}$\tablenotemark{a}} &
\colhead{$m_{0.7}$\tablenotemark{b}} &
\colhead{$m_{i}$\tablenotemark{c}} &
\colhead{SFR\tablenotemark{d}} &
\colhead{Fit\tablenotemark{e}} &
\colhead{$r_{e,0}$\tablenotemark{f}} & 
\colhead{$r^C_{1/2}$\tablenotemark{g}} &
\colhead{$r^T_{1/2}$\tablenotemark{h}} & 
\colhead{a/b\tablenotemark{i}} }
\startdata
1  &  0000-263-C07  &  ---  & Summ\tablenotemark{j}  & 25.90 & 25.27 & 25.28 &
  ---   & d & 0.08 & 0.20 & 0.20 & 1.58 \nl
2  &  0000-263-C09  & 3.428 & Summ  & 25.09 & 24.45 & 24.41 & 12 (40) & d & 0.10 & 0.20 & 0.20 & 1.53 \nl
3  &  0000-263-C10  &  ---  & F606W & 24.98 & 24.66 & 24.66 &   ---   & e & 0.09 & 0.10 & 0.10 & 1.18 \nl
4  &  0000-263-C11  & 3.150 & Summ  & 25.78 & 25.06 & 25.07 &  6 (17) & e & 0.13 & 0.20 & 0.20 & 1.13 \nl
5  &  0000-263-C12  &  ---  & Summ  & 26.76 & 25.56 & 25.31 &   ---   & e &  --- & 0.30 & 0.40 & 2.72 \nl
6  &  0000-263-C13  &  ---  & Summ  & 26.37 & 25.41 & 25.40 &   ---   & e & 0.25 & 0.25 & 0.25 & 2.39 \nl
7  &  0000-263-C14a & 3.281 & Summ  & 26.32 & 25.45 & 25.51 &  4 (13) & d & 0.47 & 0.20 & 0.25 & 1.75 \nl
7  &  0000-263-C14b & 3.281 & Summ  & 26.14 & 25.43 & 25.48 &  4 (13) & d & 0.29 & 0.20 & 0.20 & 1.94 \nl
8  &  0000-263-C20  &  ---  & Summ  & 24.52 & 24.29 & 24.34 &   ---   & p &  --- & 0.10 & 0.10 & 1.06 \nl
9  &  0000-263-C24  &  ---  & Summ  & 24.84 & 24.62 & 24.65 &   ---   & p &  --- & 0.10 & 0.10 & 1.25 \nl
10 &  0000-263-C26  &  ---  & Summ  & 26.45 & 25.67 & 25.68 &   ---   & d & 0.04 & 0.20 & 0.20 & 1.26 \nl
11 &  0000-263-C27  & 2.780 & Summ  & 26.38 & 25.03 & 24.88 &  5 (14) & e & 0.25 & 0.30 & 0.35 & 2.24 \nl
12 &  0000-263-C28  &  ---  & Summ  & 26.61 & 25.29 & ----- &   ---   & e & 1.01 & 0.30 &  --- & 2.4  \nl
13 &  0347-383-N01  &  ---  & F702W & 26.22 & 25.46 & 25.57 &   ---   & e & 0.11 & 0.20 & 0.20 & 1.31 \nl
14 &  0347-383-N02  &  ---  & F702W & 26.70 & 24.66 & 24.66 &   ---   & e & 0.14 & 0.25 & 0.25 & 1.10 \nl
15 &  0347-383-N05  & 3.243 & F702W & 24.94 & 23.99 & 23.94 & 17 (53) & d & 0.53 & 0.25 & 0.30 & 2.37 \nl
16 &  SSA22-L3-C02  &  ---  & F814W & 25.40 & 24.95 & 24.89 &   ---   & e & 0.15 & 0.15 & 0.15 & 1.18 \nl
17 &  SSA22-L3-C24  &  ---  & F814W & 25.36 & 23.86 & 23.35 &   ---   & e & 0.43 & 0.35 & 0.50 & 2.90 \nl
18 &  SSA22-CW-C12a & 3.201 & F814W & 27.01 & 25.05 & 24.80 &  7 (23) & e & 0.52 & 0.45 & 0.50 & 1.55 \nl
19 &  SSA22-CW-C12b & 3.201 & F814W & 26.57 & 25.39 & 24.80 &  7 (23) & e & 0.20 & 0.30 & 0.50 & 1.49 \nl
20 &  SSA22-CW-C16  &  ---  & F814W & 26.5  & 26.0  & 25.7  &   ---   & e & 0.17 & 0.25 & 0.30 & 1.38 \nl
\enddata
\tablenotetext{a}{Magnitude in a $0.2$ arcsec radius aperture, on the $AB$ scale}
\tablenotetext{b}{Magnitude in a $0.7$ arcsec radius aperture, on the $AB$ scale}
\tablenotetext{c}{Isophotal magnitude, on the $AB$ scale}
\tablenotetext{d}{Star formation rates from isophotal magnitudes in units of 
$h^{-2}_{50}\, $M$_{\odot}$ yr$^{-1}$ for $q_0=0.5$ (0.05)} 
\tablenotetext{e}{Function which better models the radial profile: d=$r^{1/4}$;
e=exp; p=point-source}
\tablenotetext{f}{Fitted scale length: $r_e$ for $r^{1/4}$; $r_0$ for exponential, in arcsec}
\tablenotetext{g}{Core half-light radius in arcsec, from direct photometry}
\tablenotetext{h}{Total half-light radius in arcsec, from direct photometry}
\tablenotetext{i}{Axial ratio from the isophotal aperture}
\tablenotetext{j}{Summ=F606W+F814W}
\end{deluxetable}
\newpage
\figcaption{The morphologies of the $z>3$ galaxies. The galaxies are at the
center of the circle. The scale is 0.1 arcsec/pixel, and each box is 10 
arcsec in size. The images clearly show
the compact ``core'' and the low surface brightness nebulosities which often 
surround it. Although we do find a few cases of relatively diffuse systems and
systems with multiple ``cores'', the overall dispersion of morphologies among
the $z>3$ population seems significantly narrower than that found in the 
star-forming galaxies observed at later cosmological epochs.}
\figcaption{The radial profiles of the $z>3$ galaxies. The horizontal axes are
radial separation in arcsec. Vertical ones are surface brightness in
mag/arcsec$^2$. The horizontal dot-long
dashed lines are the $1\, \sigma$ surface brightness fluctuations of the
frames in mag/arcsec$^2$. The error bars are the $1\, \sigma$ rms fluctuations
along the isophotes. The dot lines are the fitted $r^{1/4}$ profiles, the
dashed lines are the exponentials. The presence of a peaked central light
concentration surrounded by a more diffuse and irregular light distribution
can be observed in the radial profiles.}
\newpage
\begin{figure}
\figurenum{1}
\epsscale{0.8}
\plotone{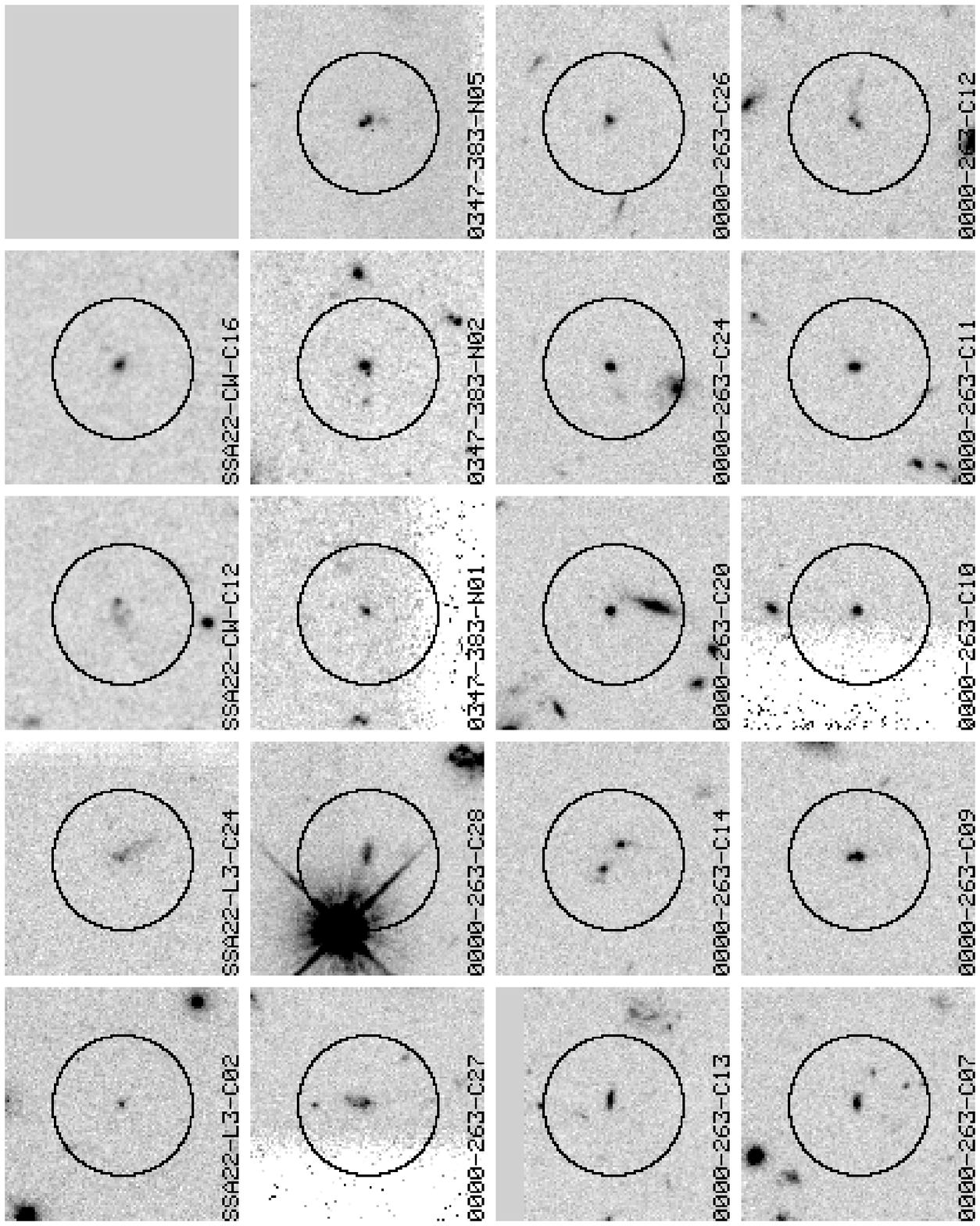}
\caption{The morphologies of the $z>3$ galaxies. The galaxies are at the
center of the circle. The scale is 0.1 arcsec/pixel, and each box is 10 
arcsec in size. The images clearly show
the compact ``core'' and the low surface brightness nebulosities which often 
surround it. Although we do find a few cases of relatively diffuse systems and
systems with multiple ``cores'', the overall dispersion of morphologies among
the $z>3$ population seems significantly narrower than that found in the 
star-forming galaxies observed at later cosmological epochs.}
\end{figure}
\newpage
\begin{figure}
\epsscale{1.1}
\figurenum{2}
\plotone{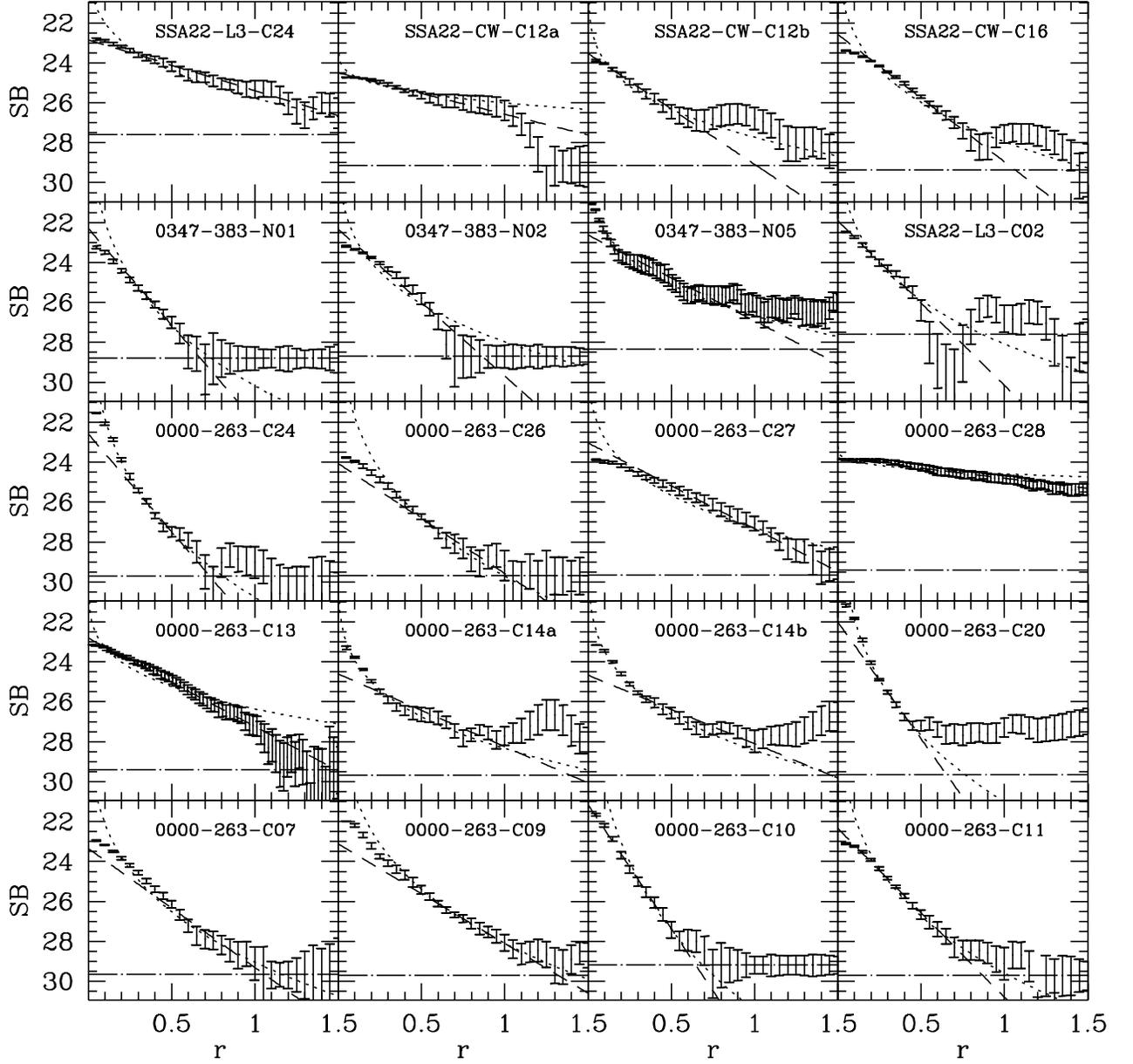}
\caption{The radial profiles of the $z>3$ galaxies. The horizontal axes are
radial separation in arcsec. Vertical ones are surface brightness in
mag/arcsec$^2$. The horizontal dot-long
dashed lines are the $1\, \sigma$ surface brightness fluctuations of the
frames in mag/arcsec$^2$. The error bars are the $1\, \sigma$ rms fluctuations
along the isophotes. The dot lines are the fitted $r^{1/4}$ profiles, the
dashed lines are the exponentials. The presence of a peaked central light
concentration surrounded by a more diffuse and irregular light distribution
can be observed in the radial profiles.}
\end{figure}
\end{document}